\documentstyle[preprint,aps,eqsecnum,epsf,floats]{revtex}
\tighten
\input epsf
\def\bfl{{\bf l}}
\def\bull{\vrule height .9ex width .8ex depth -.1ex}
\def\MeV{{\rm MeV}}

\def\Tr{{\rm Tr\,}}
\def\nrcpt{NR\raise.4ex\hbox{$\chi$}PT\ }
\def\ket#1{\vert#1\rangle}
\def\bra#1{\langle#1\vert}
\def\ltap{\ \raise.3ex\hbox{$<$\kern-.75em\lower1ex\hbox{$\sim$}}\ }
\def\gtap{\ \raise.3ex\hbox{$>$\kern-.75em\lower1ex\hbox{$\sim$}}\ }

\def\CA{{\cal A}}
\def\CC{{\cal C}}
\def\CD{{\cal D}}

\def\CL{{\cal L}}
\def\CO{{\cal O}}
\def\CZ{{\cal Z}}

\def\pds{{\it PDS}\ }

\def\bfq{{\bf q}}

\def\bfp{{\bf p}}
\def\bfpp{{\bf p '}}

\def\bfx{{\bf x}}

\def\dfx{{\rm d}^4 x\,}
\def\bfy{{\bf y}}

\def\dfy{{\rm d}^4 y\,}

\def\frac#1#2{{\textstyle{#1\over#2}}}
\def\darr#1{\raise1.5ex\hbox{$\leftrightarrow$}\mkern-16.5mu #1}
\def\){\right)}
\def\({\left(}
\def\]{\right]}
\def\[{\left[}
\def\si{{}^1\kern-.14em S_0}
\def\siii{{}^3\kern-.14em S_1}
\def\diii{{}^3\kern-.14em D_1}

\def\MeV{{\rm\ MeV}}
\def\CA{{\cal A}}

\newcommand{\eqn}[1]{\label{eq:#1}}
\newcommand{\refeq}[1]{(\ref{eq:#1})}
\newcommand{\eq}{eq.~\refeq}
\newcommand{\eqs}[2]{eqs.~(\ref{eq:#1}-\ref{eq:#2})}

\newcommand{\beq}{\begin{eqnarray}}
\newcommand{\eeq}{\end{eqnarray}}

\begin{document}

\preprint{\vbox{
\hbox{DOE/ER/40561-2-INT98}
\hbox{NT@UW-98-04}
\hbox{CALT-68-2170} }}
\bigskip
\bigskip
\title{A Perturbative Calculation of the Electromagnetic Form Factors of the
Deuteron}
\author{David B. Kaplan}  \address{ Institute for Nuclear Theory,
University of Washington, Seattle, WA 98915   \\  {\tt
dbkaplan@phys.washington.edu} }  \author{Martin  J. Savage}  \address{
Department of Physics, University of Washington,  Seattle, WA 98915
\\ {\tt savage@phys.washington.edu} }  \author{Mark B. Wise} \address{
California Institute of Technology, Pasadena, CA 91125  \\  {\tt
wise@theory.caltech.edu} }  \maketitle
\begin{abstract}
Making use of the effective field theory expansion recently
developed by the authors, we compute the electromagnetic form
factors of the deuteron analytically to next-to-leading order
(NLO).  The computation is rather simple, and involves
calculating several Feynman diagrams, using dimensional
regularization. The results agree well with data and indicate
 that the expansion is converging. They
do not suffer from any ambiguities arising from off-shell
versus on-shell amplitudes.

\end{abstract}

\vskip 2in

\leftline{April 16, 1998}
\vfill\eject

\section{Introduction}
\label{sec:1}

The techniques introduced by the authors in refs. \cite{KSW} put
the study of low energy two-nucleon interactions  on the same
footing as chiral perturbation theory in the mesonic and single nucleon sectors
 \cite{ManRev}.  In particular, there is a systematic
low momentum expansion, such that at any given order one need only
calculate a finite number of Feynman diagrams to arrive at an
analytic result. The procedure is superior in several ways to
the conventional technique of solving the Schr\"odinger equation
with a potential constructed to fit the scattering data: (i) There is a well
defined expansion parameter, and one can estimate errors at any
given order in the expansion; (ii) it is straightforward to incorporate
relativistic and inelastic effects within the expansion; (iii)
analytic results allow one to see quite simply the relative importance
of short- and long-distance physics to a given process; (iv) there
is no ambiguity concerning off-shell matrix elements when calculating
physical processes; (v) at low orders in the expansion,  the number
of free parameters to be fit to the data is few, and the same
parameters are used in all processes.  The results at lower orders in the
expansion are therefore
very constrained.

Until now, the techniques of ref. \cite{KSW} have only been applied to
reproducing
scattering phase shifts.  While a necessary first step, fitting
the  phase shifts does not seriously  test the method,
as the low energy  phase shifts can be well fit by rather simple functions of
few parameters.
What is needed are calculations of dynamical processes
that involve the same interactions as are fit to the $NN$ phase shifts.  The
obvious ones to
consider are
$NN\to NN\gamma$, $np\to d\gamma$, parity and isospin violation in $NN$
processes,
 $pp\to de^+\nu$,
 and the deuteron electromagnetic form factors.
In this paper we present the perturbative calculation of the  deuteron
electromagnetic form factors at NLO.  This subject has been addressed
previously in the context of effective field theory in refs. \cite{Bira,Park},
although using a somewhat different formalism and involved numerical, as
opposed to analytical, calculations.
We preface the calculation with a
brief review of our expansion, and a discussion of the deuteron.
After identifying the graphs contributing the electromagnetic form factors, we
show explicitly
that there is no ambiguity arising from the fact that the nucleons
in a deuteron are not on their mass shell, even though the couplings
in the effective theory are fit to $NN$ scattering data.   We conclude with a
discussion of features that will appear in the NNLO (next-to-next-to
leading order) calculation of the form factors.

It is not unreasonable to ask why it is worth pursuing an effective field
theory description of the deuteron since effective range 
theory \cite{bethea,betheb}, can be used to predict many of its properties\cite{erica,ericlot}. For some quantities, like the deuteron charge
radius, effective range theory is remarkably precise.
The primary motivation is to make a clear connection to QCD and therefore
enable systematic calculations to be performed, even for processes where
effective range theory is not applicable. Furthermore we expect that, even
in cases where effective range theory is very accurate, the effective field
theory approach will surpass this level of precision if pursued to higher
orders.

In this paper the application of the effective field theory expansion of
ref. \cite{KSW} to processes involving the deuteron is developed. For 
definiteness we focus on the electromagnetic form factors of the deuteron.
However, it is straightforward to use the methods developed here for other
quantities, like the cross sections for $np \rightarrow d\gamma$ and
$\gamma d \rightarrow \gamma d$.

At NLO the predictions of the effective field theory expansion for the
electromagnetic form factors of the deuteron are not as accurate as those
of effective range theory. However at NNLO they should reach the precision of
effective range theory. Furthermore, the effective field theory approach is
a systematic one at some order it will include physical effects beyond those
that are encorporated into effective range theory.

\section{Effective field theory for $NN$ interactions}
\label{sec:2}

In order to compute the electromagnetic form factors of the deuteron, we must
consider the possible interactions between nucleons, pions and photons. In an
effective
field theory, these interactions take the form of local operators, constrained
only by
the symmetries of QCD and QED.  In this section we discuss the form of the
operators that occur to
the order that we will be working, and then  turn to the issue of power
counting,
which allows a consistent expansion of the form factors.

\subsection{Interactions}
\label{sec:2a}
Terms in the effective Lagrangian describing the interactions
 between nucleons, pions
and photons can be classified by the number of nucleon fields that appear in
them.  It is convenient to write
\begin{equation}
{\cal L} = {\cal L}_0 + {\cal L}_1 + {\cal L}_2 + \ldots,
\end{equation}
where ${\cal L}_n$ contains $n$-body nucleon operators.

${\cal L}_0$ is constructed from the photon field $A^\mu = (A^0, {\bf A})$ and
the pion fields $\Pi$; it does not contain any nucleon fields.
 The pion fields are incorporated
in a special unitary matrix,
\begin{equation}
\Sigma = \exp {2i\Pi\over f},\qquad \Pi = \left(\begin{array}{cc}
\pi^0/\sqrt{2} & \pi^+\\ \pi^- & -\pi^0/\sqrt{2}\end{array} \right),
\end{equation}
where $f=132\ \MeV$ is the pion decay constant.  $\Sigma$
transforms under the global $SU(2)_L \times SU(2)_R$  and $U(1)_{em}$ gauge
symmetries as
\begin{equation}
\Sigma \rightarrow L\Sigma R^\dagger, \qquad \Sigma \rightarrow e^{i\alpha
Q_{em}} \Sigma
e^{-i\alpha Q_{em}}
\end{equation}
where $L\in SU(2)_L$, $R\in SU(2)_R$ and $Q_{em}$ is the charge matrix,
\begin{equation}
Q_{em} = \left(\begin{array}{cc}
1 & 0\\
0 & 0\end{array}\right).
\end{equation}

The part of the Lagrange density with no nucleon fields is
\begin{eqnarray}
{\cal L}_0 = {1\over 2} ({\bf E}^2 - {\bf B}^2) + {f^2\over 8} \,\Tr D_\mu
\Sigma
D^\mu \Sigma^\dagger
+ {f^2\over 4} \omega \,\Tr\, m_q (\Sigma + \Sigma^\dagger) + \ldots .
\end{eqnarray}
The ellipsis denotes operators with more covariant derivatives $D_\mu$,
insertions of the quark mass matrix, $m_q = {\rm diag} (m_u, m_d)$, or factors
of the
electric and magnetic fields.  Acting on $\Sigma$ the covariant derivative is
\begin{equation}
D_\mu \Sigma = \partial_\mu \Sigma + ie [Q_{em},\Sigma] A_\mu,
\end{equation}
The parameter $\omega$ has dimensions of mass and $m_\pi^2 = \omega (m_u +
m_d)$.

When describing pion-nucleon interactions it is convenient to introduce the
field $\xi = \exp i \Pi/f = \sqrt{\Sigma}$.  Under $SU(2)_L \times SU(2)_R$
transformations
\begin{equation}
\xi \rightarrow L\xi U^\dagger = U\xi R^\dagger,
\end{equation}
where $U$ is a complicated nonlinear function of $L,R$ and the pion fields
themselves.  Since $U$ depends on the pion fields it has spacetime dependence.
The nucleon fields are introduced as a doublet of spin $1/2$ fields
\begin{equation}
N = \left({p\atop n}\right),
\end{equation}
that transforms under chiral $SU(2)_L \times SU(2)_R$ symmetry as $N
\rightarrow UN$ and under $U(1)$ gauge transformations as $N \rightarrow
e^{i\alpha Q_{em}} N$.  Acting on nucleon fields the covariant derivative is
\begin{equation}
D_\mu N = (\partial_\mu + V_\mu + ie Q_{em}A_\mu )N,
\end{equation}
where
\begin{eqnarray}
V_\mu = {1\over 2} (\xi D_\mu \xi^\dagger + \xi^\dagger D_\mu \xi)\ .
\end{eqnarray}
The covariant derivative of $N$ transforms in the same way as $N$ under
$SU(2)_L \times SU(2)_R$ transformations (i.e., $D_\mu N \rightarrow U D_\mu
N$) and under $U(1)$ gauge transformations (i.e., $D_\mu N \rightarrow
e^{i\alpha Q_{em}} D_\mu N$).

The one-body terms in the Lagrange density are
\begin{eqnarray}
{\cal L}_1 &&= N^\dagger \left(i D_0 + {{\bf D}^2\over 2M}\right) N 
+ {ig_A\over
2} N^\dagger {\bbox \sigma} \cdot (\xi {\bf D} \xi^\dagger - \xi^{\dagger} {\bf
D} \xi)
N\nonumber \\
&&+ {e\over 2M} N^\dagger \left( \kappa_0 + {\kappa_1\over 2} [\xi^\dagger
\tau^3 \xi + \xi \tau^3 \xi^\dagger]\right) {\bbox \sigma} \cdot {\bf B} N +
\ldots
\qquad .
\eqn{lagone}
\end{eqnarray}
To the order to which we are working 
$\kappa_0 = {1\over 2} (\kappa_p + \kappa_n)$ and
$\kappa_1 = {1\over 2} (\kappa_p - \kappa_n)$
are isoscalar and isovector nucleon magnetic moments in nuclear magnetons, with
\begin{eqnarray}
\kappa_p & = & 2.79285\ ,\qquad
\kappa_n = - 1.91304 \ ,
\eqn{kapdef}
\end{eqnarray}
at tree-level.
At higher orders there will be contributions to \eq{kapdef} from pion loop graphs
\cite{jenmag}. 
The isoscalar magnetic moment $\kappa_0$ receives leading corrections of the form 
$m_\pi^2\log (m_\pi^2/\Lambda_\chi^2)$,  suppressed by two powers of the pion mass.
In contrast, the isovector magnetic moment $\kappa_1$
receives leading corrections of the form $m_\pi$, suppressed by only one power of
the pion mass.
The ellipsis in \eq{lagone} denotes higher order terms that do not contribute at
the order we are working.

Finally it remains to consider the  two body operators.  Some of these were
discussed in
refs. \cite{KSW}; however, since we will be computing electromagnetic form
factors of
the deuteron there are additional considerations that didn't arise in the NLO
calculation of nucleon phase shifts.

 First we will consider the two-body operators involving nucleons alone, then
we will look at those containing a photon; to the order we will be working, we
need not consider two-body operators involving pion fields.
In the spin triplet channel, there is one $ NN $ contact interaction with no
derivatives or insertions of the quark mass matrix, corresponding to a diagonal
transition  $\siii\to\siii$;  the coefficient of this operator is taken to be
$C_0$. There is an additional contact interaction involving no derivatives and
 one insertion of the
quark mass matrix, with coefficient $D_2$; it can be distinguished from the
$C_0$
 interaction by its chiral properties.  There are five contact interactions
involving two gradients, corresponding to diagonal transitions in the
$ {}^3\kern-.14em S_1,
{}^3\kern-.14em P_0,  {}^3\kern-.14em P_1$, and $ {}^3\kern-.14em P_2$ partial
waves, as well as an off-diagonal  $\siii-\diii$ transition.  Only the first
and the last of these are relevant for the deuteron;  furthermore, at NLO we
can ignore the $\siii-\diii$ transition interaction.  Thus the only $\nabla^2$
two-body contact interaction we will consider is $\siii\to \siii$, and has
coupling $C_2$.  Therefore for a $\siii\to\siii$ scattering process, where the
incoming nucleons have momenta $\bfp_1$, $\bfp_2$ and polarization $i$, and
scatter into states with
momenta $\bfp'_1$, $\bfp_2'$ and polarization $j$, the Born amplitude  arising
from the contact
interactions is
\beq
i\CA = -i\delta_{ij}\[ C_0  + D_2 m_\pi^2 + {C_2\over 8}\((\bfp_1 - \bfp_2)^2
+
(\bfp'_1-\bfp'_2)^2\)\]\ .
\eeq
The form of the $C_2$ amplitude is fixed by Lorentz invariance (which is
equivalent to Galilean invariance to the order we work), and by the
normalization we used in ref. \cite{KSW}, where in the center of mass frame,
where we defined the amplitude to be $-iC_2 p^2$, $p\equiv \vert
\bfp_i\vert=\vert \bfp'_i\vert$ \footnote{The couplings $C_0$, $D_2$ and $C_2$
are the same couplings that appear as $C^{(\siii)}_0$,  $ D^{(\siii)}_2$, and
  $C^{(\siii)}_2$ in refs. \cite{KSW};   we drop the $\siii$ designation here
as there can be no confusion with analogous couplings in the $\si$ channel.}.
As discussed in appendix~\ref{sec:7}, while one can construct a two-body
contact interaction with one factor of $\partial_0$ instead of two gradients,
for any $S$-matrix element (including those involving the deuteron)  one can
use the equations of motion to eliminate time derivatives for gradients. Thus
no independent $\partial_0$ contact interaction needs to be introduced.

Including gauge fields introduces several two-body contributions to the
electromagnetic current.  Firstly, the $C_2$ interaction described above
becomes gauged.  Secondly there are two new two-body magnetic moment type
interactions. In order to write $\CL_2$ compactly we define the matrix $P_i$
which projects onto the $\siii$ state,
\beq
\label{project}
P_i \equiv {1\over \sqrt{8}} \sigma_2\sigma_i\tau_2\ , \qquad \Tr P_i^\dagger
P_j
=-\Tr P_i^\dagger P_j^T
={1\over 2}\delta_{ij}\ ,
\eeq
where the $\sigma$ matrices act on the nucleon spin indices, while the $\tau$
matrices act on isospin indices.
Then the two-body Lagrangian may be written as
\beq
\eqn{contact}
\CL_2 &=& -\(C_0+ D_2 \omega\,\Tr m_q\) (N^T P_i N)^\dagger(N^T P_i N)
\nonumber\\
 & + & {C_2\over 8}
\left[(N^T P_i N)^\dagger
\left(N^T \left[ P_i \overrightarrow {\bf D}^2 +\overleftarrow {\bf D}^2 P_i
    - 2 \overleftarrow {\bf D} P_i \overrightarrow {\bf D} \right] N\right)
 +  h.c.\right]
\nonumber\\
&&+ e L_2  \[ (N^T P_i N)^\dagger (N^T P_i {\bbox \sigma}\cdot{\bf B} N) +
h.c.\]
+\ldots\ ,
\eqn{lagtwo}
\eeq
where
the ellipsis refers to contact interactions irrelevant for the deuteron
channel, or of higher order than we will be considering.  The new coupling
$L_2$ corresponds to an interaction that did not enter the
calculation of $NN$ scattering, but which affects the deuteron magnetic form
factor.
As written, \eq{lagtwo} is not chirally invariant, which can be remedied by an
appropriate
insertion of the $\xi$ fields;  however, since the two-body operators with
pions do not contribute at NLO, we omit them.

\subsection{Power counting}
\label{sec:2b}

We begin by summarizing the results of refs. \cite{KSW}. The
starting point is the effective Lagrangian for nucleons, pions
and photons introduced in the previous section.  
The part of the Lagrangian describing purely
mesonic interactions, as well as interactions between mesons and
a single baryon, is the conventional chiral Lagrangian.
In addition there are local
interactions corresponding to short distance interactions between two nucleons.
These
contact interactions are expanded in powers of derivatives and insertions of
the quark
mass matrix, $m_q$. (Isospin violation from the difference between
the up and down quark masses is neglected. Consequently
insertions of $m_q$ are equivalent to factors of $m_{\pi}^2$.)
The lowest dimension operator is  a four fermion contact interaction;
there are two independent operators of this form, corresponding to the $\si$
and $\siii$
channels.
The next lowest dimension two-body  operators involve a factor of $p^2$, where
$p$ is the momentum of one of the nucleons in the center of mass frame, or a
factor of
$m_\pi^2$. There are seven independent $p^2$ operators corresponding to
diagonal matrix
elements in the $\{{}^1\kern-.14em S_0,  {}^1\kern-.14em P_1,  {}^3\kern-.14em
S_1,
{}^3\kern-.14em P_0,  {}^3\kern-.14em P_1, {}^3\kern-.14em P_2\}$ channels, as
well as a
$\siii-\diii$ mixing term; there are two independent $m_\pi^2$ operators
corresponding to the
$\si$ and
$\siii$ channels.  At higher powers of derivatives, the number of contact
interactions quickly grows.

Central to effective field theory is a power counting scheme which allows one
to calculate
consistently to any given order in the low energy expansion. A main point in
refs. \cite{KSW}
was to develop the $PDS$  subtraction scheme which allows one to readily
identify the order
of any particular Feynman graph.   The scheme involves computing loop diagrams
using dimensional
regularization, and then subtracting off the poles in dimensions $D\le 4$,
which correspond to
logarithmic or power-law divergences.  A typical integral in this scheme is
\beq
\openup3\jot
I_n&\equiv& i(\mu/2)^{4-D} \int {{\rm d}^D q\over (2\pi)^D}\, {\bf q}^{2n}
\({ E/2 + q_0 -{\bf q}^2/ 2M + i\varepsilon}\)^{-1}\({E/2 - q_0
-{\bf q}^2/2M + i\varepsilon} \)^{-1}
\nonumber\\
&=& (\mu/2)^{4-D} \int {{{\rm d}}^{(D-1)}{\bf  q}\over (2\pi)^{(D-1)}}\,
{\bf q}^{2n} \({E  -{\bf q}^2/M + i\varepsilon}\)^{-1}
\nonumber\\
&=& -M (ME)^n (-ME-i\varepsilon)^{(D-3)/2 } \Gamma\({3-D\over 2}\)
{(\mu/2)^{4-D}\over  (4\pi)^{(D-1)/2}}\nonumber\\
&\mathop{\longrightarrow}\limits_{\scriptstyle{D\to 4}}^{PDS}\ & - (ME)^n
\left({M\over 4\pi}\right)
(\mu  -\sqrt{-ME-i\varepsilon})\ .
\eqn{pds}
\eeq
The last step includes the finite subtraction mandated in the $PDS$ scheme.
The parameter $\mu$ is the renormalization scale and physical observables are
independent of it.
In fact,  one may set $\mu$ to zero and recover the usual minimal subtraction
scheme ($MS$)
with $\mu=0$ if one wishes \footnote{In the $MS$ scheme with $\mu=0$
one must
first integrate out the pion to avoid factors that diverge  as
$\log (m_\pi^2/\mu^2)$.}. However, a change in $\mu$ must be compensated by the
renormalization
group flow of the couplings in the theory. Therefore, what is a weak coupling
at one
value of $\mu$ can be strong at another, which effects how one defines the
power counting
scheme.

Rapid scaling with $\mu$ is only an issue for two body operators,
and then only for those affected by the large scattering lengths in the $\si$
and $\siii$ channels.
Consider  a four nucleon  contact interaction
connecting angular momentum states $L$ and $L'$,
where conservation of angular momentum and parity requires
$\vert L-L'\vert$ to equal zero or two.
We assume that the operator involves  $m$ insertions of the
quark mass matrix, and $2d \equiv (L+L'+2n)$ spatial
gradients,
and has a coefficient $\CC^{L,L'}_{m,n}$.  By examining the coupled
renormalization group
equations in the \pds scheme,  one can determine that  these couplings scale as
\beq
 \CC^{L,L'}_{m,n}(\mu)\sim \cases{
 \mu^{-(m+n+1)} &$L,\ L' \in \{\si,\,\siii,\,\diii\}$;\cr
 \mu^{ 0} & otherwise.\cr}
\eqn{cscale}
\eeq
in the region $1/a \ll \mu \ll \Lambda_{NN}$, where \beq
\Lambda_{NN} = {8\pi f^2\over g_A^2 M}\simeq 300\ \MeV.
\eeq
Here $M$ is the nucleon mass, $g_A=1.25$ is the axial current coupling and
$f=132\,\MeV$ is the pion decay constant. Thus in the deuteron channel,
$C_0\sim \mu^{-1}$, while $C_2$ and $D_2$ scale as $\mu^{-2}$.  Extending the
analysis to include photons, we find $L_2\sim \mu^{-2}$ as well.

The  coefficients of the
four-nucleon contact terms that have explicit factors of
the electric field ${\bf E}$ or the magnetic field ${\bf B}$ scale similarly to
those in \eq{cscale}, counting gauge fields as derivatives.
For example, the  $L_2$ operator in \eq{contact} counts as a  two-derivative,
$L=L'=0$ operator,
and its coefficient
scales as $L_{2} \sim \mu^{-2}$.
The rapid scaling of the operators contributing to $S
$-wave processes is what
makes our
expansion different than the one  proposed by Weinberg \cite{Weinberg}.

Armed with the above results, we are able to arrive at a particularly simple
set of rules
for determining
the order of a graph.  Choosing the scale   $\mu\sim p \sim m_\pi\sim Q$ we
perform an
expansion in $Q$, where
\begin{enumerate}
\item Each nucleon or pion propagator scales as $Q^{-2}$;
\item Each loop integration $\int {\rm d}^4q$ scales as $Q^5$;
\item A gradient at a vertex contributes $Q^1$, while each time derivative
scales as $Q^2$;
\item An insertion of the quark mass matrix $m_q$ at a vertex counts as $Q^2$;
\item The coefficient of the contact interactions scale according to
\eq{cscale}.
\end{enumerate}
The first three rules follow simply from the scaling of four momenta $q_\mu$
appropriate to the nonrelativistic regime. Explicitly, $Mq_0\sim \bfq^2\sim
Q^2$. The fourth rule is familiar from conventional
 chiral perturbation theory, $m_q\sim m_\pi^2\sim Q^2$.
The procedure for calculating physical quantities of interest
 is to write down the most general effective field theory consistent with
gauge invariance, chiral symmetry and Lorentz invariance \footnote{Relativistic
corrections are accounted for as perturbations according to the above power
counting rules,
and at the order we work the theory only appears Galilean invariant.
The procedure for dealing with relativistic corrections perturbatively requires
distinguishing between potential and radiation pions at NNLO, as discussed in
\cite{KSW}.},
and then
compute the desired matrix element to a given order in the $Q$ expansion,
following the above rules.
Note that according to the power counting rules, a loop with two propagators
entails a factor of $Q$, while
the coefficient of the lowest order $NN$ contact
interaction ($\CC^{LL'}_{d,m}$ with $L=L'=d=m=0$,
defined to be $C_0$) scales
as $1/Q$;  thus any graph may be dressed by an infinite bubble chain with $C_0$
interactions
without changing
the order of the graph.

\bigskip\bigskip
\section{The deuteron form factors}
\label{sec:3}

A deuteron with four-momentum $p^\mu$ and polarization vector $\epsilon^\mu$ is
described by the state $\ket{\bfp ,\epsilon}$, where the polarization vector
satisfies $p_\mu\epsilon^\mu=0$.  An orthonormal basis of
polarization vectors $\epsilon_i^\mu$
satisfies
\beq
p_\mu\epsilon_i^\mu=0\ ,\qquad \epsilon_{i\mu}^*\epsilon^\mu_j =- \delta_{ij}\
,\qquad \sum_{i=1}^3\epsilon^{*\mu}_i\epsilon^\nu_i = {p^\mu p^\nu\over
M_d^2}-\eta^{\mu\nu}\ ,
\eeq
where $M_d$ is the deuteron mass.
It is convenient to choose the basis polarization vectors so that in
the deuteron rest frame $\epsilon_i^\mu=\delta^{\mu}_i$.
Deuteron states with these polarizations are denoted by
$\ket{\bfp,i}$ (i.e., $\ket{\bfp,i} \equiv \ket{\bfp,\epsilon^\mu_i}$)
and satisfy the normalization condition
\beq
\langle\bfpp,j\vert\bfp,i\rangle ={p^0\over M_d}(2\pi)^3\delta^3(\bfp-\bfpp)
\delta_{ij}\ .
\eeq

In terms of these states and to leading order in the nonrelativistic expansion,
the matrix element of the electromagnetic current is
\beq
\label{defn}
\bra{\bfpp,j} J^0_{em}\ket{\bfp,i} &=& e \[  F_C(q^2) \delta_{ij} + {1\over 2
M_d^2}F_{\cal Q}
(q^2)\(\bfq_i\bfq_j-{1\over 3}\bfq^2 \delta_{ij}\)\]\ ,\nonumber\\
\bra{\bfpp,j} {\bf J}^k_{em}\ket{\bfp,i}&=& {e\over 2 M_d} \[ F_C(q^2)
\delta_{ij}(\bfp+\bfpp)^k + F_M(q^2)\(\delta_j^k\bfq_i - \delta_i^k\bfq_j\)
\right.\nonumber\\
&&\qquad\quad \left.+{1\over 2M_d^2} F_{\cal Q}(q^2) \(\bfq_i\bfq_j-{1\over
3}\bfq^2\delta_{ij}\)(\bfp+\bfpp)^k\]\ ,
\eqn{formfactors}
\eeq
where  ${\bf q}={\bf p}^{\prime}-{\bf p}$ and $q=|{\bf q}|$.
These dimensionless form factors are normalized such that \cite{Zuilhof}
\beq
F_C(0) &=& 1\ ,\nonumber\\
{e\over 2 M_d}F_M(0) &=& \mu_{M}\ ,\nonumber\\
{1 \over M_d^2} F_{\cal Q}(0) &=& \mu_{\cal Q}\ ,
\eqn{normalization}
\eeq
where $\mu_M= 0.85741 (e/2M)$ is the deuteron magnetic moment, and $\mu_{\cal
Q}=0.2859\,{\rm fm}^2$ is the
deuteron quadrupole moment.

As shown in appendix~\ref{sec:6}, the form factors are readily calculated by
computing in perturbation theory the irreducible two-point function  $\Sigma$,
and the irreducible three-point function $\Gamma^\mu$. In the present context,
``irreducible'' means  the sum of graphs which do not fall apart when cut at
any $C_0$ vertex. The matrix element of the electromagnetic current is then
given by the exact relation
\beq
\bra{\bfpp, j} J^\mu_{em} \ket{\bfp, i}=
 i \[  {\Gamma^\mu_{ij}(\overline E,\overline E', \bfq)\over
{\rm d}\Sigma(\overline E)/{\rm d}E}\]_{\overline E,\, \overline E'\to -B}\ ,
\eqn{current}
\eeq
where
 $B$ is the deuteron binding energy and $\overline E$ is the energy
 of the incoming two nucleon state in the center of mass frame,
 \beq
\overline E\equiv  E - {\bfp^2\over 4M} +\ldots\ ,\qquad E\equiv (p^0-2M),
\eqn{ebar}
\eeq
 where the ellipsis refers to relativistic corrections to the
 energy-momentum relation.  $\overline E'$ is the analogous quantity for the
outgoing nucleon pair.
 By Lorentz invariance, $\Sigma$ and $\Gamma^\mu$ can
only depend on the
energy and momentum in this combination.

\begin{figure}[t]
\centerline{\epsfxsize=6.5in \epsfbox{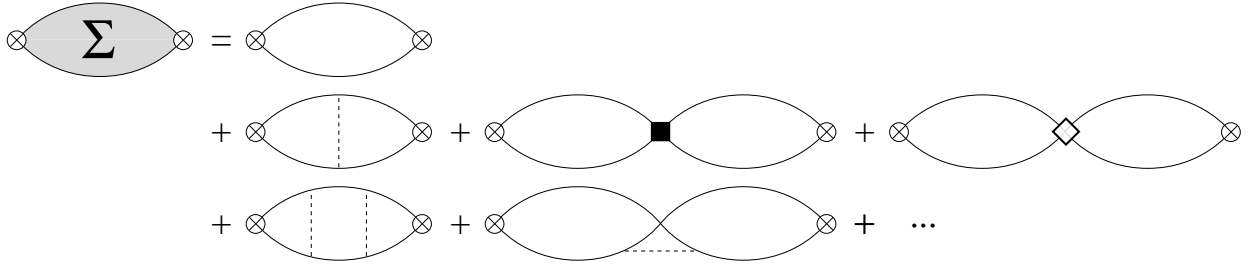}}
\noindent
\vskip.2in
\caption{\it The perturbative expansion of $\Sigma$. The first row has the
leading $O(Q)$ result; $\otimes$ represents an insertion of the interpolating
field defined in \eq{interp}.  The second row has the complete subleading
$O(Q^2)$ contribution,  where $\bull$ and $\diamond$  denote the $C_2$ and
$D_2$ interactions respectively.
The third row shows a couple of $O(Q^3)$  NNLO contributions,
which we do not calculate here: the exchange of two potential pions, and the
dressing
of $C_0$ (the pointlike $NN$ vertex)
by a radiation pion.}
\label{sigma}
\vskip .2in
\end{figure}

We can now expand the relation \eq{current} in perturbation theory and
determine the form factors by comparing the result with \eq{formfactors}.  The
two-point function has the graphical expansion shown in Fig.~\ref{sigma}, where
the $\otimes$ vertices represent the insertion of an interpolating field ${\cal
D}_i$ with the quantum numbers of a deuteron with polarization $i$.  We take
${\cal D}_i$ to be
\beq
{\cal D}_i \equiv N^T P_i N ,
\eqn{interp}
\eeq
 where $P_i$ is the projection defined in eq. (\ref{project}). The form factor
one calculates does not  depend on the  particular choice for $\CD_i$, so long
as it is used consistently.

By examining the graphs and using the power counting outlined in the previous
section, one sees that $\Sigma$ begins at order $Q^1$ --- the leading graph has
two nucleon  propagators and one loop.  At subleading order, $O(Q^2)$,  there
are three two-loop graphs, one involving the exchange of a potential pion
(which has a derivative coupling),
 one with an insertion of the $C_2 (\tensor{{\bf D}})^2$
two-body operator, and one with an insertion of the $D_2m_\pi^2$ two-body
operator. Recall that with renormalization scale $\mu \sim Q$ the coefficients
$C_2$ and $D_2$
are $O(Q^{-2})$.  At
$O(Q^3)$ there are a host of diagrams, including the exchange of two potential
pions, or one radiative pion, as well as $\bfp^4$ relativistic corrections to
the nucleon propagator, etc.  We have calculated $\Sigma$ to $O(Q^2)$, and
the results are presented in appendix~\ref{sec:6}.

\subsection{The NLO computation of the electric form factors}
\label{sec:3b}

\begin{figure}[t]
\centerline{\epsfxsize=6.5in \epsfbox{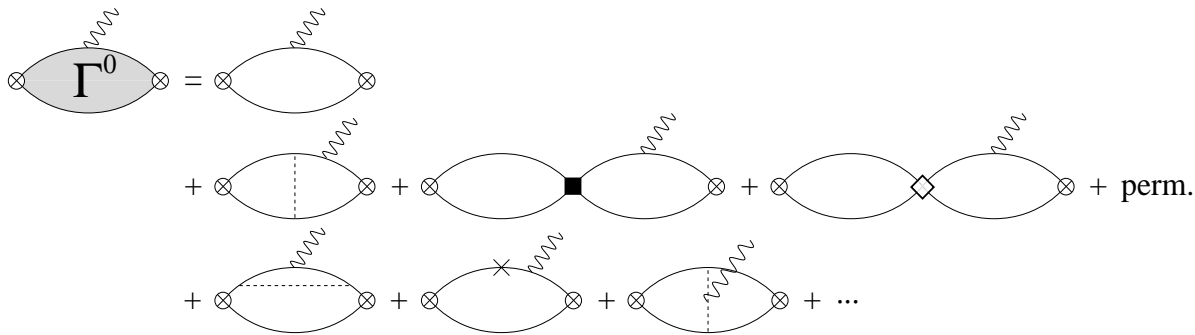}}
\noindent
\caption{\it The expansion of $\Gamma^0$. In all of these graphs, the
photon corresponds to $A_0$ with the minimal coupling to the proton
propagator, arising from the gauged nucleon kinetic energy term.
The graph in the first row is the leading $O(Q^{-1})$ contribution,
$\Gamma^0_{(-1)}$.
 The second row are the  subleading graphs at   $O(Q^0)$, summing to give
$\Gamma^0_{(0)}$.  In the third row are several graphs contributing at the
$O(Q^1)$: a dressing
of the photon-nucleon vertex, a relativistic
correction to the nucleon propagator, and an exchange current contribution. }
\label{gamma0}
\vskip .2in
\end{figure}

To compute the electric form factors $F_C$ and $F_{\cal Q}$ we need to
calculate the three-point function $\Gamma^0_{ij}$, which is  expanded
graphically in Fig.~\ref{gamma0}.  The results for  both the leading
$O(Q^{-1})$
and subleading $O(Q^0)$ contributions are presented in appendix~\ref{sec:6}.
Once $\Gamma^0$ is computed in the $Q$-expansion,
the electric form factors can be determined by expanding \eq{current} as
\beq
\bra{\bfpp, j\,} J^0_{em} \ket{\bfp, i}=
 i \[  {\Gamma^0_{(-1)}\over {\rm d}\Sigma_{(1)}/{\rm d}\overline E }\]
+i\[  {\Gamma^0_{(0)}\({\rm d}\Sigma_{(1)}/{\rm d}\overline E\) -
\Gamma^0_{(-1)}
\({\rm d}\Sigma_{(2)}/{\rm d}\overline E\)\over \({\rm d}\Sigma_{(1)}/{\rm
d}\overline E\)^2}\]
+ O(Q^2)\ ,
\eqn{current0}
\eeq
where $\Gamma^0_{(n)}$, $\Sigma_{(n)}$ denote the $O(Q^n)$ contribution to
$\Gamma^0$ and $\Sigma$ respectively. We have suppressed the $\bfq$ dependence
of $\Gamma^0$, and its polarization indices. Furthermore
everything is evaluated on-shell, $\overline E=\overline E' = -B$. Since
${\rm d}/{\rm d}\overline E\sim O(Q^{-2})$, the first bracket in
\eq{current0} is $O(Q^0)$, the second bracket is $O(Q^1)$, etc.  Therefore,
taking into account the explicit factors of $q$ in the definition of the form
factors, \eq{formfactors}, we see the electric form factors have a $Q$
expansion of the form
\beq
F_C &=& F_C^{(0)} + F_C^{(1)} + O(Q^2)\ ,\nonumber\\
F_{\cal Q} &=& F_{\cal Q}^{(-2)} + F_{\cal Q}^{(-1)} + O(Q^0)\ .
\eeq
where $F^{(n)}\sim O(Q^n)$.

Using eqs. (\ref{eq:current0},\ref{eq:zb},\ref{eq:gam0m1}) gives our leading
result for the electric form
factors,
\beq
F_C^{(0)} ( q^2)  &=&
{4\gamma\over {q}} \tan^{-1}\left( {{q}\over 4\gamma}\right)\ ,
\nonumber\\
F_{\cal Q}^{(-2)} ({q^2}) &=& 0\  ,
\eqn{fe00}
\eeq
where we have defined
\beq
\gamma=\sqrt{MB}\ .
\eeq

The subleading  form factors are extracted from
 eqs. (\ref{eq:current0},\ref{eq:zb},\ref{eq:green0}),
 and presented in terms of a Feynman parameter
integral.  The electric monopole form factor is given by
\beq
F_C^{(1)} ({q^2}) & = &
- C_2(\mu)\ {M\gamma (\mu-\gamma)^2\over 2\pi}
\left[ 1- {4\gamma\over {q}}\tan^{-1} \left( {{q}\over 4\gamma}\right) \right]
\nonumber\\
& - &
{g_A^2 M m_\pi^2 \gamma\over 2 \pi f^2 {q}}
\left[ {2\over (m_\pi+ 2\gamma)  } \tan^{-1} \left( {{q}\over 4\gamma}\right)
- \int_0^1 dx\ {1\over x   \Delta }
\tan^{-1} \beta
\right]\ .
\eqn{fe01}
\eeq
where we have defined the functions
\beq
\Delta(x)= \sqrt{\gamma^2 +x(1-x)q^2/4}\ ,\qquad
\beta(x)    =  { {q} x \over 2 (\gamma + m_\pi + \Delta  )}
\ .
\eqn{deltadef}
\eeq
The operator with coefficient $D_2$ does not contribute to these observables.
 Because of the running of $C_2$,
the above expression is independent of $\mu$ to the order we are working
\cite{KSW}.
{}From \eqs{fe00}{fe01} we determine the charge radius of the deuteron
to NLO,
\begin{eqnarray}
\langle r^2\rangle^{LO} & = & {1\over 8\gamma^2}
\nonumber\\
\langle r^2\rangle^{NLO} & = & C_2(\mu)\ {M (\mu-\gamma)^2\over 16\pi\gamma}
\ +\
{g_A^2 M m_\pi^2 (3 m_\pi+10\gamma)\over 96 \pi f^2 \gamma (m_\pi + 2
\gamma)^3}
\ .
\eqn{chargeradius}
\end{eqnarray}
A comparison with the experimental value is given in \S\ref{sec:4}.

At NLO, the electric quadrupole form factor is given by
\beq
{F_{\cal Q}^{(-1)} ({q^2}) \over M_d^2}&=&
{3 g_A^2 M\gamma \over  \pi  16 \pi f^2 {q}^3}
\int_0^1 dx\
{1\over x \beta^4 \Delta }
\nonumber\\
& &
\times\left(
\left[ 3 {q}^2 x^2 (1+\beta^2 )^2 - 24 {q} m_\pi \beta   x (1+\beta^2 )
+ 16 m_\pi^2 \beta  ^2 (3+\beta^2 )\right] \tan^{-1} \beta
\right.
\nonumber\\
& & \left.
\ +\beta   \left[ -48 m_\pi^2\beta^2 + 8 m_\pi {q} x \beta    (3+2\beta^2 )
- {q}^2 x^2 (3+5\beta^2 )
\right]
\right)
\eqn{formsub}
\eeq
{}From this expression one can extract the quadrupole moment to first
nonvanishing order:
\beq
\mu_{\cal Q}^{LO} & = & 0
 \ ,\qquad
\mu_{\cal Q}^{NLO} =
{g_A^2 M (6\gamma^2 + 9 \gamma m_\pi + 4 m_\pi^2)\over 30\pi f^2 (m_\pi+2
\gamma)^3}
\ .
\eqn{quadmoment}
\eeq
A comparison with the experimental value is given in \S\ref{sec:4}.

\subsection{The NLO computation of the magnetic form factor}
\label{sec:3c}

\begin{figure}[t]
\centerline{\epsfxsize=6.5in \epsfbox{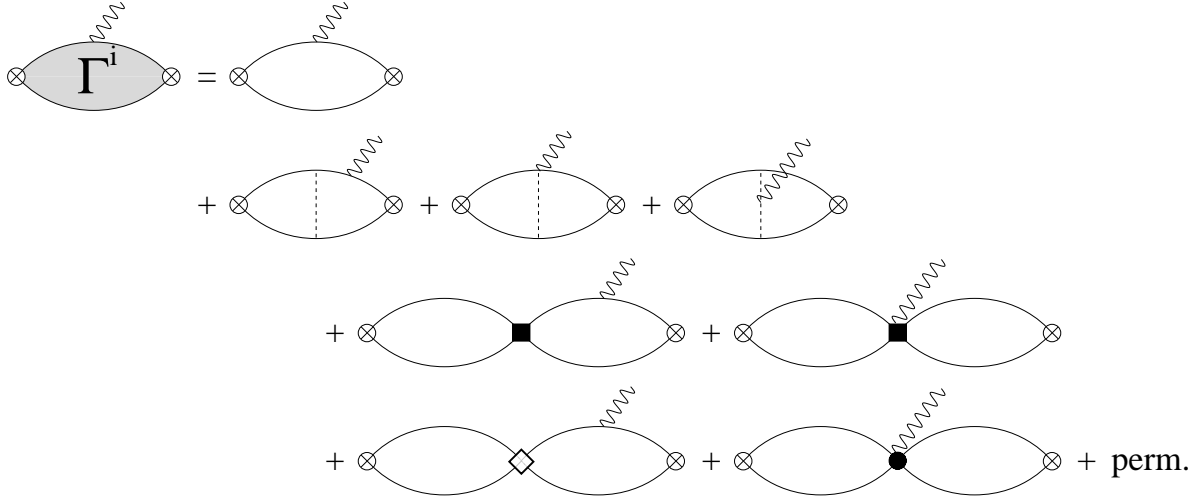}}
\noindent
\caption{\it
The expansion of $\Gamma^i$, where the photon corresponds to the
vector potential ${\bf A^i}$. The coupling of the photon to the nucleon lines
represents the entire one-body current from
$\CL_1$, \eq{lagone}, including the magnetic moment contribution. The first
graph is the LO contribution at $O(Q^0)$,
while the remaining graphs are the NLO contributions at $O(Q^1)$.  The  photon
couplings arise through any of the
operators in $\CL_0$, $\CL_1$ or $\CL_2$. We specifically distinguish the
$C_2$, $D_2$ and $L_2$ vertices by the
symbols $\bull$, $\diamond$, and $\bullet$ respectively. }
\label{gammai}
\vskip .2in
\end{figure}

In order to calculate the magnetic form factor of the deuteron, we need the
matrix element of the spatial current  $\bra{\bfpp, k\,} {\bf J}^i_{em}
\ket{\bfp, j}$.  This entails computing $\Gamma^i$, using the coupling of the
spatial component of the gauge field, ${\bf A}_i$, discussed in \S\ref{sec:2a}.
  The
expansion of $\Gamma^i$ in Feynman graphs is shown to subleading order in
Fig~(\ref{gammai}). Following our power counting rules,  $\Gamma^i$ begins at
$O(Q^0)$, and so an expansion analogous to \eq{current0} for the matrix element
of  ${\bf J}^i_{em} $ implies that the magnetic form factor has the expansion
\beq
F_M = F_M^{(0)} + F_M^{(1)} + O(Q^2)\ .
\eeq
Our task in computing $F_M$ is greatly simplified by recognizing from
\eq{formfactors} that we need only pick out contributions with spin structure
anti-symmetric in
the deuteron polarization vectors.
It is straightforward to check that none of the graphs shown in
Fig.~(\ref{gammai}) contribute to $F_M$ when the photon coupling arises from
any of the operators
$N^\dagger {\bf D}^2 N$,
$g_A  N^\dagger {\bbox \sigma} \cdot (\xi {\bf D} \xi^\dagger - \xi^{\dagger}
{\bf
D} \xi)N$,
 $Tr\left[ D_\mu\Sigma D^\mu\Sigma^\dagger \right]$ in \eq{lagone},  or the
four-nucleon
operator with coefficient $C_2$ in \eq{lagtwo}.
At LO, only  the photon coupling via the isosinglet  nucleon magnetic moment
one-body operator
contributes,
\beq
{e\over 2 M}\kappa_0 N^\dagger{\bbox\sigma}\cdot {\bf B} N = {\mu_p+\mu_n\over
2} N^\dagger{\bbox\sigma}\cdot {\bf B} N\ ,
\eqn{mag}
\eeq
and we find
\beq
{e F_M^{(0)}(q^2)\over 2 M_d} & = &
{e\over M}\kappa_0\,F_C^{(0)}({q}^2)\ =\
(\mu_p+\mu_n)
{4\gamma\over {q}} \tan^{-1} \left( {{q}\over 4\gamma}\right)
\ .
\eeq
For the deuteron magnetic moment this gives $\mu_M^{LO} = (\mu_p+\mu_n)$,
simply the sum of the neutron and proton magnetic moments.

At next order, $Q^1$, there are contributions to $F_M$ arising from coupling
the photon via \eq{mag}, along with insertions of the $C_2$ operator or one
pion exchange; there is also a contribution from the two-body current arising
from the operator in \eq{lagtwo} whose coefficient is  $L_{2}$.  We find that
there are no pion
exchange current contributions at this order, nor any two-body current
contribution from the $C_2$ operator in \eq{lagtwo}.  With the exception of the
two-body contribution involving an explicit factor of ${\bf B}$
(see \eq{contact}),
 all the graphs  contributing are all proportional to those
giving rise to the electric form factors in Fig.~(\ref{gamma0}).  Therefore to
this order we can express the
magnetic form factor in terms of the
electric form factors and a single new coupling constant.  We find
\beq
{e F_M^{(1)}(q^2)\over 2 M_d} & = & (\mu_p+\mu_n)\,\left(
F_C^{(1)}({q}^2)
+{{q}^2 \over 12 M_d^2}F_{\cal Q}^{(-1)}({q}^2)
\right)
+ e L_2{\gamma\over \pi} (\mu -\gamma)^2
\ ,
\eeq
and the deuteron magnetic moment is given by
\beq
\mu_M^{LO} &=& \mu_p+\mu_n\ ,\nonumber\\
\mu_M^{NLO} &=& { e L_2} {\gamma\over \pi} (\mu -\gamma)^2\ .
\eqn{magmoment}
\eeq
where $L_2$ depends on
the renormalization scale $\mu$ in such a way that $\mu_M^{(1)}$ is
$\mu$-independent.
A comparison with the experimental value is given in the next section.

\subsection{Effective range theory}
\label{sec:3d}
In effective range theory the electromagnetic form factors are assumed to
be dominated by the asymptotic S-wave deuteron wave function,
\begin{eqnarray}
  \psi^{({\rm ER})}({\bf r}) & = & \sqrt{{\gamma \over 4 \pi
 (1- \gamma r_0) }} {e^{-\gamma r} \over r}
  \ \ \ .
\end{eqnarray}
Assuming the small r part of the deuteron wave function is only important
for establishing the normalization condition, $F_C(0)=1$, the prediction of
effective range theory for the form factor $F_C(q^2)$ follows from the
 Fourier transform of $ |\psi^{(ER)}({\bf r})|^2$,
\beq
F^{(ER)}_C(q^2)=
1\ +\ \left({1 \over 1- \gamma r_0} \right)
\left(-1 + 
{4 \gamma \over q} \tan^{-1}\left({q \over 4 \gamma}\right) \right)
\ \ \ .
\eeq
This yields the charge radius,
\begin{eqnarray}
   \langle r^2\rangle^{\rm ER}  & = & {1\over 8\gamma^2} {1\over 1 - \gamma
     r_0}
   \nonumber\\
    & = & {1\over 8\gamma^2}\left[ 1 + \gamma r_0\ + \gamma^2 r_0^2\ +\
      ...\right]
    \ \ \ .
\end{eqnarray}
It is instructive to compare the effective range theory prediction for the
charge radius with that from effective field theory. In effective field theory
at NLO the effective range is ($r_0=0$ at LO),
\begin{eqnarray}
r_0 & = &   C_2(\mu)\ {M (\mu-\gamma)^2\over 2\pi }
\ +\ { g_A^2 M \over 4 \pi f^2} \left( 1 - {8 \over 3}{\gamma\over m_\pi} +
  2{\gamma^2\over m_\pi^2} \right)
\ \ \ .
\end{eqnarray}
Using this it is straightforward to show that the $\gamma$ expansion of the NLO
effective field theory charge radius,
\begin{eqnarray}
  \langle r^2\rangle & = & {1\over 8\gamma^2}
  \left[
    1
    \ +\ C_2(\mu)\ {M \gamma (\mu-\gamma)^2\over 2\pi }
    \ +\ {g_A^2 M \gamma m_\pi^2 (3 m_\pi+10\gamma)\over 12 \pi f^2 (m_\pi + 2
      \gamma)^3}
    \right]
\nonumber\\
 & = &
 {1\over 8\gamma^2}
  \left[
    1
    \ +\ C_2(\mu)\ {M \gamma (\mu-\gamma)^2\over 2\pi }
    \ +\ { g_A^2 M \gamma \over 4 \pi f^2} \left( 1 - {8 \over 3}
{\gamma\over m_\pi} + 4
      {\gamma^2\over m_\pi^2}\ +\ ...\right) 
    \right]
\ \ \ ,
\end{eqnarray}
and the $\gamma$ expansion of effective range theory agree to order $\gamma^2$
at linear order in $r_0$.

Effective range theory predicts the matter radius, $r_{\rm m}$,
with remarkable precision.
Using $r_0=1.75\ {\rm fm}$ effective range theory yields
$r^{(ER)}_{\rm m}=1.98\ {\rm  fm}$.
The most recent measurement of the
deuteron charge radius is  $r_{\rm ch} = 2.1303\pm 0.0066\ {\rm fm}$
from which the matter radius is found to be
$r_{\rm m} = 1.9685\pm 0.0049\ {\rm fm}$\cite{buch}.
(In effective
field theory the effects that distinguish between the matter and charge
radius don't arise until NNLO.)
The numerical success of the prediction of
effective range theory for the matter radius suggests that the most important
higher order terms in effective field theory are those that arise from
iterating the NLO potential from $C_2$ and one-pion exchange.
However, from
the effective field theory perspective this cannot be justified since there
are new local operators that will contribute at the same order.

Effective range theory can also be used to predict the magnetic form factor
and it gives,
\beq
{e F_M^{(ER)}(q^2) \over 2 M_d}=(\mu_p+\mu_n)F_C^{(ER)}(q^2)
\eeq
In the following section $F_C^{(ER)}(q^2)$ and $F_M^{(ER)}(q^2)$ are compared
with experimental data.

\section{Comparison with data}
\label{sec:4}
We now compare the analytic results of our effective field theory perturbative 
expansion for the deuteron form factors with
experimental data. We have evaluated these expressions at the same
renormalization point $\mu=m_\pi$ used in refs. \cite{KSW} and have used the
same value
\beq
C_2(m_\pi) = 9.91\ {\rm fm}^4
\eeq
derived from a fit to the $NN$ scattering phase shifts in the spin triplet
channel.  The values of $C_0$ and $D_2$ do not enter our expressions
explicitly, but they do enter indirectly through the constraint on the
two-point function that the deuteron pole occurs at the correct binding energy,
\eq{zb}.
Given $C_2$ from the $NN$ phase shift analysis, we have no new parameters at
through NLO for fitting the electric form factors.  As we have seen, for the
magnetic form factor, a single new parameter, $L_2$, enters at NLO.

\begin{table}[tbp] \centering
\caption{Electromagnetic properties of the deuteron \label{table1}}
\begin{tabular}{ccccc}
& LO & NLO & (LO+NLO) & Experiment \cite{buch,Ericson}  \\ \hline
RMS charge radius (fm) & 1.53 & 0.36  & 1.89 & 2.1303(66) \\
\noalign{\smallskip}
Magnetic moment (N.M.) & 0.88 & -0.02 (fit) & 0.86 (fit)& 0.85741
\\
\noalign{\smallskip}
Quadrupole moment (${\rm fm}^2$)  & - & 0.40 & 0.40& 0.2859(3) \\
\end{tabular}
\end{table}
We first consider that static moments of the deuteron, at $q^2=0$.
We have analytic formulas
for the charge radius, the quadrupole moment, and the magnetic moment in
eqs.~(\ref{eq:chargeradius}), (\ref{eq:quadmoment}), and  (\ref{eq:magmoment})
respectively.
A comparison of these values to experiment is given in
Table~\ref{table1}.
The charge radius shows a rapid convergence to the
measured value, which is encouraging.
The LO calculation is expected to be within $\sim 30\%$ of the experimental
value, while the NLO calculation is expected to be within $\sim 10\%$.
It is clear from Table~\ref{table1} that this expectation is fulfilled.
When the NNLO calculation is performed we expect that the result is within
$\sim 3\%$ of the experimental value.
The magnetic moment agrees well with
experiment at LO, and then is fit to the experimental value
at NLO by choosing the strength $L_2$ of the two-body magnetic operator
appropriately.
The LO prediction for the magnetic moment is much closer to the experimental
value (within $\sim3\%$) than naively expected from the power counting.  
The quadrupole moment vanishes at LO, and
the NLO value of $0.40\ {\rm fm}^2$ 
is off by $\sim 40\%$, as expected from the power counting.
It would be useful to compute the NNLO contribution to $\mu_{\cal Q}$ to see if
it exhibits
the same convergence as the charge radius.
The idea of including pions perturbatively has been used previously to estimate
the deuteron quadrupole moment\cite{wong}, obtaining a value of  $0.40\ {\rm fm}^2$.
More interesting is that iterated potential pion exchange
reproduces the deuteron
quadrupole moment\cite{ericlot} reasonably well.
This suggests that  contributions to the quadrupole moment 
from higher order counterterms are small compared to additional 
insertions of potential
pion exchange.  This smallness is not something that arises naturally in the
effective field theory.  It is also interesting that state-of-the-art
nuclear calculations of the quadrupole moment \cite{mccalc}
($\sim 0.270\ {\rm fm}^2$) are systematically lower than the
experimental value by $\sim 7\% $.  This strongly suggests that dynamics
beyond potential interactions are required, something that 
effective field theory provides a systematic way to include.

\begin{figure}[t]
\centerline{\epsfxsize=3.5in \epsfbox{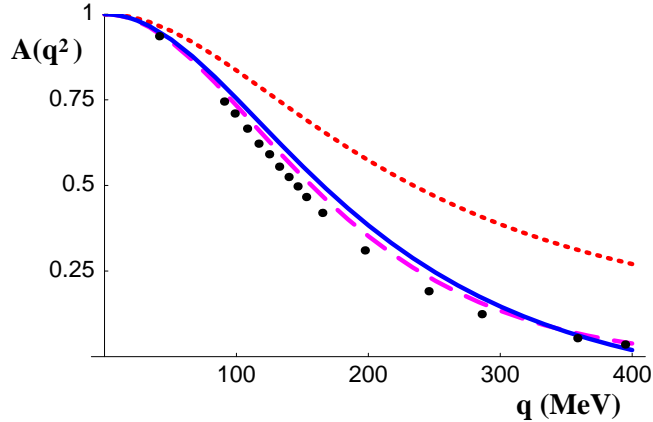}}
\noindent
\caption{\it
  A plot of $A(q^2)$ vs. $q$ in MeV for elastic electron-deuteron
scattering.
The dotted curve shows the result of the LO calculation, while the
solid curve is the NLO prediction.
There are no free parameters at this order.
The dashed curve shows the result of effective range theory.
}
\label{aplot}
\vskip .2in
\end{figure}

\begin{figure}[t]
\centerline{\epsfxsize=3.5in \epsfbox{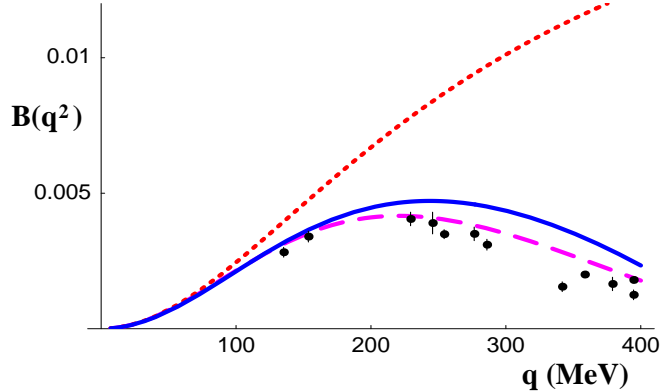}}
\noindent
\caption{\it
    A plot of $B(q^2)$ vs. $q$ in MeV for elastic electron-deuteron
scattering.
The dotted curve shows the result of the LO calculation, while the
solid curve is the NLO prediction.
There is one free parameter at this order, $L_2$, which is fixed to correctly
reproduce the deuteron magnetic moment.
The dashed curve shows the result of effective range theory.
}
\label{bplot}
\vskip .2in
\end{figure}

Of greater interest is the comparison of the form factors over a range of
$q^2$, as we should be able to see at what momentum the expansion begins to
fail;  our naive estimate is that the expansion is in powers of $q/
2\Lambda_{NN}\sim q/(600\,\MeV)$.
 The differential cross section for elastic electron-deuteron scattering is
given  by
\beq
{{\rm d}\sigma\over {\rm d}\Omega} = {{\rm
d}\sigma\over {\rm d}\Omega}\biggl\vert_{\rm Mott}\biggr.\[A(q^2) +
B(q^2)\tan^2\theta/2\]\ ,
\eeq
where $A$ and $B$ are related to the form factors  \cite{Zuilhof}:
\beq
A &=& F_C^2 + {2\over 3}\eta  F_M^2 + {8\over 9}\eta^2 F_{\cal Q}^2\
,\nonumber\\
B &=& {4\over 3}\eta(1+\eta) F_M^2\ ,
\eqn{abf}
\eeq
with $\eta\equiv -(p- p')^2/(4 M_d^2)\simeq q^2/4M_d^2$.  In order to compare
with data, we take our
analytic results for the form factors and expand the expression \eq{abf} in
powers of $Q$, where $\eta\sim O(Q^2)$
\beq
A&=& \[ \(F_C^{(0)}\)^2\] + \[2  F_C^{(0)} F_C^{(1)}\] + O(Q^2)\ ,\nonumber\\
B&=& \[{4\over 3}\eta \(F_M^{(0)}\)^2\] + \[{8\over 3}\eta F_M^{(0)}
F_M^{(1)}\] + O(Q^4)\ .
\eeq
We see that to the order we are working, $A$ is sensitive only to the electric
form factor $F_C$, while $B$ depends only on the magnetic form factor $F_M$.  A
comparison of $A$ and $B$ with experimental data in
Figs.~(\ref{aplot},\ref{bplot}) shows that our expansion is quite successful,
and converging rapidly, in the kinematic regime where it is expected  to work.
The
data for Fig.~\ref{aplot} was taken from ref. \cite{Simon}, and the error bars
are smaller than the size of the points; the data for Fig.~\ref{bplot} comes
from
refs.~\cite{Simon,Grosstete,Benaksas,Ganichot}.

It is evident from Figs.~(\ref{aplot},\ref{bplot})
that the NLO effective field theory calculation of the
deuteron form factors in not as accurate as what effective range theory gives. The validity of effective range theory over such a wide range of momentum
occurs because of the smallness of the shape parameter, $r_1$.
In the effective field theory expansion, the coefficients in the effective
range expansion themselves have perturbative expansions.
However, ultimately when carried out to higher orders
the effective field theory calculations will be more precise than the effective
range calculations.
This is because the effective field theory correctly describes the strong
interactions, which effective range theory only approximates.

\section{Conclusions}
\label{sec:5}

We have demonstrated that one can compute properties of the two nucleon
system to surprising accuracy simply by calculating several
Feynman
diagrams. The technique for doing this was introduced in refs. \cite{KSW}
where it had been shown how to work at NLO for $NN$ phase shifts in both
spin
singlet and triplet channels.  While encouraging, those results were not
definitive as  the NLO calculation required three free parameters in
both
spin channels.  The true test of the theory has been presented in this
paper
with the computation of the electromagnetic form factors of the deuteron
---
by using the parameters fit to  scattering data, we are able to
reproduce very well at NLO both the electric and magnetic form factors
in
elastic $e-d$ scattering up to momentum transfers
$q^2=(400\ \MeV)^2 = 4.1\ {\rm fm}^{-2}$.
Since our results  are analytic, it is straightforward to analyze what
features in the data are due to short versus long distance physics.
 A central feature of our expansion --- that pion
exchange is perturbative --- is supported by the success of our fit to
the form factors.

One feature of our results which is especially encouraging is the
evidence
that the expansion is converging rapidly.  This is apparent in the
improvement of the fits to $e-d$ scattering data in going from LO to
NLO,
improvements in the
static moments of the deuteron.  The RMS charge radius presented in
Table~\ref{table1} deviates from  the experimental value by
$\sim 30\% $
at leading order, but only $\sim 10\% $
at next-to-leading order.  The magnetic moment was off by $\sim 3\%$
at leading order, and exact at next-to-leading order, due to the contribution
of a new operator. At NLO the results of effective field theory for the
electric and magnetic form factors of the deuteron, $F_C(q^2)$ and $F_M(q^2)$,
are not as accurate as those from
effective range theory. However at NNLO the effective field theory approach
should reach (or even surpass) the precision of effective range theory. Furthermore, the methods developed in this paper can be used to 
make predictions for other properties of the deuteron, including 
those for which effective range theory is not applicable.

Since the NLO result for the quadrupole form factor is the first
nonvanishing
term in its expansion, it is expected to work less well.  At the level we
are
working, the quadrupole form factor does not contribute to $e-d$
scattering,
however,  we can compare the quadrupole moment with experiment, and it is $\sim
40\%$
too large.  We expect  this error to be substantially reduced in the
NNLO
calculation, which includes among other things the exchange of two
potential
pions, and short distance $\siii-\diii$ transitions.  In general, it would be
interesting to compare NNLO results for
all
of the form factors. Other effects that enter at this order are relativistic
corrections,
radiation pions, and nucleon form factors.

There remain a number of NLO calculations to be done in the two nucleon
system, and we are optimistic about their success.  Extending this
procedure
to the three body system and beyond remains a fascinating challenge \cite{BP}.


\vskip2in
\centerline{\bf ACKNOWLEDGMENTS}
\bigskip
We thank G. Rupak and N. Shoresh for showing us a simple way to
compute
some of the Feynman graphs we encountered.
This work supported in part by the U.S. Dept. of Energy under
Grants No. DOE-ER-40561,  DE-FG03-97ER4014, and DE-FG03-92-ER40701.

\appendix

\section{The graphical expansion of the matrix element of $J^\mu_{em}$}
\label{sec:6}

\subsection{Irreducible Green functions}
\label{sec:6a}
In this appendix we derive \eq{current} which is central to our calculation of
the deuteron electromagnetic form factors.  We begin with the interpolating
field
defined in the text,
\beq
{\cal D}_i \equiv  N^T P_i N\ ,
\eqn{interp2}
\eeq
where $P_i$ is the projection defined in eq. (\ref{project}). The full
propagator $G$ is defined as the time ordered product of two of these
$\CD$ fields:
\beq
G(\overline E) \,\delta_{ij}= \int \dfx
e^{-i(Et-\bfp\cdot\bfx)}\,\bra{0} {\rm T}\[\CD_i^{\dagger}(x)\CD_j(0)\]\ket{0}
=\delta_{ij}{i\CZ(\overline E)\over  \overline E + B + i\varepsilon}\ ,
\eqn{fullprop2}
\eeq
where $B$ is the deuteron binding energy.
By Lorentz invariance, the propagator only depends on the
energy in the center of mass frame, namely
\beq
\overline E\equiv  E - {\bfp^2\over 4M} +\ldots\ ,\qquad E\equiv (p^0-2M),
\eeq
where the ellipses refers to relativistic corrections to the dispersion
relation.
 The numerator $\CZ$ in \eq{fullprop2} is assumed to be smooth near
the deuteron pole, and when evaluated at the pole gives the
wavefunction renormalization $Z$,
\beq
\CZ(-B) \equiv Z = -i \left[ {{\rm d}G^{-1}(\overline E)\over
{\rm d} E}\right]_{\overline E=-B}^{-1} .
\eeq

\begin{figure}[t]
\centerline{\epsfxsize=6.5in \epsfbox{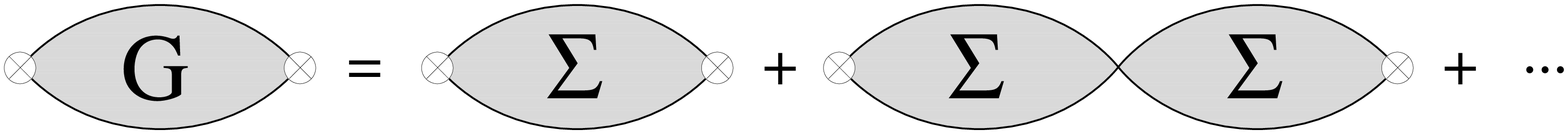}}
\vskip.2in
\noindent
\caption{\it The expansion of of the full 2-point function $G$ in terms of the
irreducible 2-point function $\Sigma$.}
\label{gfig}
\vskip .2in
\end{figure}

It is convenient to define ``irreducible'' Green functions as the sum
of graphs which
do not fall apart when the graph is cut between incoming and outgoing nucleons
at
the four-fermion vertices proportional to  $C_0$.  The irreducible 2-point
function is denoted by $\Sigma$, and has the expansion shown in
Fig.~\ref{sigma}. One can see graphically (Fig.~\ref{gfig} that the relation
between $G$ and $\Sigma$ is
\beq
G = {\Sigma\over 1+i C_0 \Sigma}\ .
\eqn{gsig}
\eeq
It follows that
\beq
\Sigma\Bigl\vert_{\overline E=-B}\Bigr. = {i\over C_0}\ ,
\qquad {1\over \Sigma^2}{{\rm d}\Sigma\over {\rm d} E}
\biggl\vert_{\overline E=-B}\biggr. = {i\over  Z}.
\eqn{zb}\eeq
In general, unphysical quantities such as $Z$, $C_0$, the
deuteron wavefunction, etc. will depend on the renormalization
scale $\mu$, while $S$-matrix elements will be $\mu$-independent.

\begin{figure}[t]
\centerline{\epsfxsize=6.5in \epsfbox{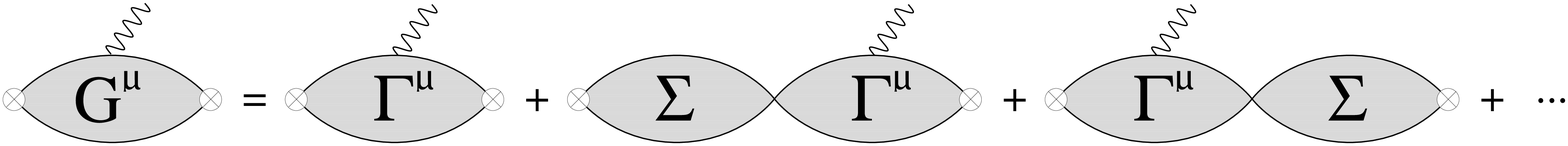}}
\noindent
\caption{\it The expansion of full three-point function $G^\mu$ in terms of the
irreducible two- and three--point functions $\Sigma$, $\Gamma^\mu$.}
\label{gmu}
\vskip .2in
\end{figure}

In order to compute the matrix element of the electromagnetic
current between two deuteron states, we first define the 3-point function
\beq
G^\mu_{ij}(\overline E,\overline E', \bfq)= \int \dfx \dfy
e^{-i(Ex^0-\bfp\cdot\bfx)}e^{i(E'y^0-\bfpp\cdot\bfy)}\,\bra{0}{\rm T}
\[\CD_i^{\dagger}(x)J_{em}^\mu(0)\CD_j(y)\]\ket{0}\
,
\eeq
where $q^\mu = (E'-E,\bfpp-\bfp)$ is the photon momentum. $G^\mu$ is
related to the desired
form factor via the LSZ formula
\beq
\bra{\bfpp, j\,} J^\mu_{em} \ket{\bfp, i}=
Z\[G^{-1}(\overline E)G^{-1}(\overline E')  G^\mu_{ij}(\overline E,\overline
E', \bfq)
\]_{\overline E,\,\overline E'\to -B}\ ,
\eqn{jme}
\eeq
where $G(\overline E)$ is defined in \eq{fullprop2}.
It is convenient to reexpress this formula in terms $\Sigma$ and the
irreducible 3-point function, which we call $\Gamma^\mu$.
It is easy to see graphically (Fig.~\ref{gmu}) that the relation between
$G^\mu$
and $\Gamma^\mu$ is
\beq
G^\mu_{ij}(\overline E,\overline E', \bfq) &=& {\Gamma^\mu_{ij}
(\overline E,\overline E', \bfq)\over \(1+i C_0\Sigma(\overline E)\)
\(1+i C_0\Sigma(\overline E')\)  }\nonumber\\
&=&{\Gamma^\mu_{ij} (\overline E,\overline E', \bfq)G(\overline E)
G(\overline E')\over \Sigma(\overline E)\Sigma(\overline E')}\ .
\eeq
Making use of this relation and eqs.~(\ref{eq:gsig}-\ref{eq:zb},\,\ref{eq:jme})
allows us to reexpress the matrix element of the current in terms of
$\Gamma^\mu$ and $\Sigma$:
\beq
\bra{\bfpp, j\,} J^\mu_{em} \ket{\bfp,i\,}=
&=&Z\[  {\Gamma^\mu_{ij}(\overline E,\overline E', \bfq)
\over \Sigma(\overline E)
\Sigma(\overline E')}\]_{\overline E,\, \overline E'\to -B}\nonumber\\
&=& i \[  {\Gamma^\mu_{ij}(\overline E,\overline E', \bfq)
\over
{\rm d}\Sigma(\overline E)/{\rm d}E}\]_{\overline E,\, \overline E'\to -B}\ .
\eqn{current2}
\eeq
It is this relation that has a simple perturbative description in terms of
Feynman graphs.

\subsection{Computing $\Sigma$}
\label{sec:6b}

We can now compute $\Sigma$ in our perturbative expansion,
writing $\Sigma$ as
\beq
\Sigma(\overline E)=\sum_{n=1}^\infty \Sigma_{(n)}(\overline E)
\eeq
where $\Sigma_{(n)}(\overline E)\sim O(Q^n)$.  The leading contribution to
$\Sigma$ is
shown in the first row of Fig.~\ref{sigma}, and is $O(Q)$ according to the
rules of
the previous section. These graphs are readily evaluated using the formula
\eq{pds},
with the result
\beq
\Sigma_{(1)}(\overline E) = -i{M\over 4\pi}\left(\mu -\sqrt{-M\overline
E-i\varepsilon}\right)\ .
\eqn{sigi}
\eeq
The subleading contribution is $O(Q^2)$ and one must compute the three graphs
shown in the second row of  Fig.~\ref{sigma}.  The result is \cite{KSW}
\beq
\Sigma_{(2)}(\overline E) &=& -i{g_A^2 M^2m_\pi^2\over 32 \pi^2
f^2}\[ i \tan^{-1}\({2\sqrt{M\overline
E}\over m_\pi}\)-{1\over 2}\ln\({m_\pi^2+4M\overline E\over
\mu^2}\)+1\]\nonumber\\
& & -i\({g_A^2\over 2 f^2}+C_2 M\overline E + D_2 m_\pi^2\)
\[\Sigma_{(1)}(\overline E)\]^2
\eqn{sigii}
\eeq
To the order we are working we truncate the expansion in \eq{ebar} to the
nonrelativistic
result,
\beq
\overline E \simeq E-{\bfp^2\over 4M}\ ;
\eeq
the first relativistic correction enters at NNLO, or $O(Q^3)$.  Other NNLO
contributions
are shown in the third row of  Fig.~\ref{sigma}, and include the exchange of
two potential
pions, or one radiative pion (see \cite{KSW} for discussion) as well as several
other graphs.

{}From \eq{current} we see that what is needed is ${\rm d}\Sigma/{\rm
d}\overline E$ evaluated
at $\overline E=-B$.  From \eqs{sigi}{sigii} we find
\beq
{{\rm d}\Sigma_{(1)}\over {\rm d}\overline E}\Bigl\vert_{\overline E=-B}\Bigr.
&=&   -i{ M^2\over 8\pi\gamma} \nonumber\\
{{\rm d}\Sigma_{(2)}\over {\rm d}\overline E}\Bigl\vert_{\overline E=-B}\Bigr.
&=&  -i{ M^3\over 16\pi^2\gamma}
\[{g_A^2\over 2 f^2}\(\gamma-\mu + {m_\pi^2\over m_\pi+2\gamma}\)
+D_2 m_\pi^2 \(\gamma-\mu\)-
C_2\gamma(\mu-\gamma)(\mu-2\gamma)\]\ ,
\eeq
where we have defined
\beq
\gamma\equiv \sqrt{MB}\ .
\eqn{gamdef}
\eeq

\subsection{Computing $\Gamma^0$}
\label{sec:6c}

The leading contribution to the matrix element of the $J^0_{em}$ current
between
deuteron states arises from the
three-point function $\Gamma^0_{(-1)}$, the first graph in Fig.~\ref{gamma0},
\beq
\Gamma^{0}_{(-1)} = -e \delta_{ij} {M^2\over 2\pi {q}}
\tan^{-1}\left( {{q}\over 4\gamma}\right)\ ,
\eqn{gam0m1}
\eeq
where $q=\vert\bfq\vert$ is the magnitude of the photon 3-momentum, and
$\gamma$
was defined above in \eq{gamdef}.

At subleading order we need to sum the diagrams in the second row
of Fig.~\ref{gamma0}. In each case, there is a minimally coupled
$A_0$ photon coupled to the proton propagator, with either an
insertion of the $C_2$ or $D_2$ contact interactions, or a single
pion exchange\footnote{One might worry that in fact there are four-nucleon
contact
interactions involving the combination a covariant time derivative $D_0$, and
hence a direct photon coupling to the $N^\dagger N^\dagger NN$ vertex.  In
fact, such an
operator may be eliminated by using the equations of motion. We
demonstrate this by explicit calculation in appendix~\ref{sec:7}.}.
We find
\beq
\Gamma^0_{(0)}  &=&
e \delta_{ij} {M^3\over 16\pi^2}
\left[
D_2(\mu) {4 m_\pi^2 (\mu-\gamma)\over {q}}
\tan^{-1} \left( {{q}\over 4\gamma}\right)
+C_2(\mu)(\mu-\gamma)
\left(  \mu-\gamma - {4\gamma^2\over q}  \tan^{-1} \left( {{q}\over
4\gamma}\right)
\right)\right.
\nonumber\\
&&+\left.
\left({g_A\over f}\right)^2
\left(
{2 (\mu-\gamma) \over {q}} \tan^{-1} \left( {{q}\over 4\gamma}\right)
\ -\
\int_0^1 dx\ {m_\pi^2\over x {q} \Delta(x)}
\tan^{-1} \left( {x {q}\over 2(\Delta(x) + \gamma +m_\pi) }\right)
\right)
\right]
\nonumber\\
&&+
e \left(  {\bfq}_i{\bfq}_j - {q}^2 \delta_{ij}\right)
{9 g_A^2 M^3\over 16\pi^2 f^2 {q}^3}
\int_0^1 dx\,\int_0^\infty dr\
{e^{-(\Delta(x)+\gamma+m_\pi)r}\over x r^3 \Delta(x)}
\left( 3 + 3 m_\pi r + m_\pi^2 r^2\right)\nonumber\\
& & \qquad
\times\left[ {2\over x {q} r}\cos\left({x {q} r\over 2}\right)
+\left( {1\over 3} - {4\over x^2 {q}^2 r^2} \right)
\sin\left({x {q} r\over 2}\right)
\right]\ ,
\eqn{green0}
\eeq
where $\Delta(x)$ is defined in \eq{deltadef}.

As discussed in the text, the calculation of the parts of $\Gamma^i$ which are
antisymmetric in the deuteron polarizations is completely analogous to the
complete calculation of $\Gamma^0$ presented here.
\section{No off-shell ambiguity --- an explicit computation}
\label{sec:7}

When working with potential models for $NN$ interactions one often faces
ambiguities about
how to continue matrix elements off-shell.  In an effective field theory
approach, there
is no such ambiguity.  All uncertainties arising in a consistent
calculation are due to
higher order operators neglected at the order one is working \cite{Politzer}.
To
illustrate this, we consider
the effect of the operator
\begin{eqnarray}
{\cal O} & = &   (N^T P_i N)^\dagger
\[ i D_0  (N^T P_i N) + \left( ({{\bf
D}^2\over 2M}N^T) P_i N +
N^T P_i ({{\bf D}^2\over 2 M}N) \right)\]\ ,
\eqn{odef}
\end{eqnarray}
where $D_\mu$ is the gauge covariant derivative\footnote{To be chirally
invariant,
 the covariant derivative should include
pion fields, but as the pion couplings do not enter to the order we are
working, we have
set them to zero.}
,
\beq
D_\mu =  \partial_\mu +ieQ_{em} A_\mu\ ,
\eeq
$Q_{em}$ being the electric charge matrix.
The operator ${\cal O}$ is not Galilean invariant but nonetheless we can
in principle consider how it enters the NLO calculation of the
deuteron form factors
via the graphs in Fig.~\ref{EqoMfig}.  However, to the order we are
working, it vanishes by
the equations of motion,
\begin{eqnarray}
\left(  iD_0 + {{\bf D}^2\over 2 M} \right) N({\bf x},t) =0\ .
\eeq
One might naively think that the equations of motion imply that the
operator $\CO$ will not
enter a calculation of $NN$ phase shifts (as the nucleons are on-shell
in that process), yet
that $\CO$ will affect deuteron matrix elements, since the nucleons are
not on-shell in a
bound state.  This would mean that a new constant enters the deuteron
calculation which
cannot be determined vis $NN$ scattering.

 However, this is reasoning is incorrect, and we now show by explicit
calculation that
operator $\CO$ does indeed vanish when considering deuteron matrix
elements.  This result
is consistent with general theorems of field theory that state that
off-shell matrix elements
are arbitrary (they can be changed by making a field redefinition) and
that the $S$-matrix elements never depend on them (even when the matrix
element is between bound states).

As an example, consider  the contribution to the deuteron
three-point function $\Gamma^0$ of the operator $\CO$ in the graphs of
Fig.~\ref{EqoMfig},
corresponding to the matrix element
\beq
\Gamma^0 = \bra{0} T\[ \CD_i^\dagger(E,{\bf 0}) \CD_j(E',\bfq)
A^0(q^0,\bfq)\]\ket{0}\ ,
\eeq
where $E'=E+q^0$.
The first graph,  Fig.~\ref{EqoMfig}(a), includes the photon-independent
part of $\CO$ and a
minimally coupled $A^0$ photon on a nucleon leg.  It is  proportional to
\begin{eqnarray}
(a) \propto
-i\int {d^{D-1} {\bf k}\over (2\pi)^{D-1}} {d^{D-1} {\bfl}\over
(2\pi)^{D-1}}\,{ E-{\bf k}^2/ M
\over
\left( E-{{\bf k}^2\over M}   \right)
\left( E-{{\bfl}^2\over M}   \right)
\left( E^\prime-{{\bfl}^2 + { ({\bfl}+\bfq)}^2\over 2M}
\right)}\
&=& 0\ ,
\eqn{grapha}
\end{eqnarray}
\begin{figure}[t]
\centerline{\epsfxsize=6.5 in \epsfbox{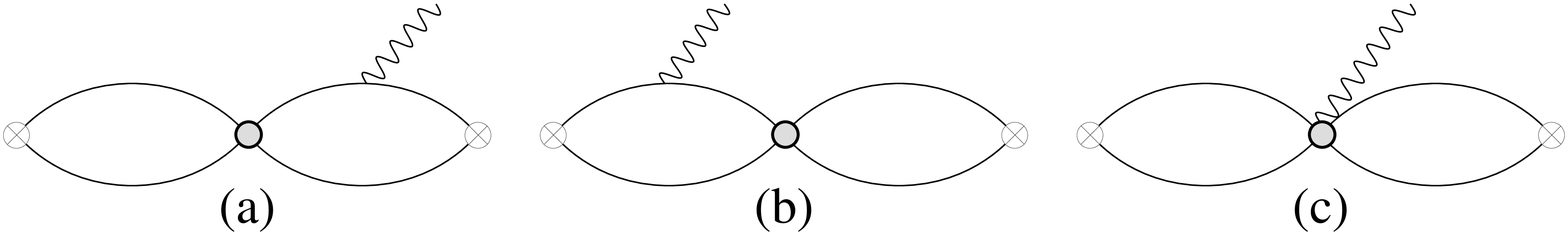}}
\noindent
\caption{\it Feynman diagrams contributing to the matrix element of
${\cal O}_1$ and ${\cal O}_2$.  The gray circle denotes an insertion of the
operator $\CO$
in \eq{odef}.}
\label{EqoMfig}
\vskip .2in
\end{figure}
where $D\to 4$ at the end of the calculation. To evaluate this integral,
we  used the fact
that the first term in the numerator cancels the first propagator,  and
that  in  dimensional regularization
\begin{eqnarray}
\int {d^{D-1} {\bf k}\over (2\pi)^{D-1}} & = & 0\ .
\end{eqnarray}

The second graph in Fig.~\ref{EqoMfig} is similar, and proportional to
\beq
(b) & \propto  &
-i\int {d^{D-1} {\bf k}\over (2\pi)^{D-1}} {d^{D-1} {\bfl}\over
(2\pi)^{D-1}}\,
{E' - \(\bf k^2 + (k + q)^2 \right)/ 2M\over
\left( E-{{\bf k}^2\over M}   \right)
\left( E^\prime-{{\bf k}^2 + {\bf (k+q)}^2\over 2M} \right)
\left( E^\prime-{{\bfl}^2 + {\bf ({\bfl}+q)}^2\over 2M}
\right)}\ ,\nonumber\\
&=&
-i\int {d^{D-1} {\bf k}\over (2\pi)^{D-1}} {d^{D-1} {\bfl}\over
(2\pi)^{D-1}}\,{1\over
\left( E-{{\bf k}^2\over M}   \right)
\left( E^\prime-{{\bfl}^2 + {\bf ({\bfl}+q)}^2\over 2M}
\right)}
\eqn{graphb}
\eeq

Finally, the third graph,  Fig.~\ref{EqoMfig}(c),  arises from the $A^0$
photon coupling in
$\CO$, and gives
\begin{eqnarray}
(c)  & \propto  &
+i\int {d^{D-1} {\bf k}\over (2\pi)^{D-1}} {d^{D-1} {\bfl}\over
(2\pi)^{D-1}}\,
{1\over \left( E-{{\bf k}^2\over M}   \right)
\left( E^\prime-{{\bfl}^2 + {\bf ({\bfl}+q)}^2\over 2M}
\right)}
\ .
\eqn{graphc}
\end{eqnarray}
It follows that  the sum of the three graphs in Fig.~\ref{EqoMfig}
vanishes, and there is
no off-shell ambiguity arising from this new operator $\CO$.  Similar remarks
hold for other operators with a single time derivative. Therefore
we can choose to
only include the two spatial derivative operators
in the Lagrangian (our interaction proportional to $C_2$) and to
eliminate the analogous operators
with a time derivative by the equations of motion --- {\it even though
we are considering nucleons
bound in a deuteron.} This result is not peculiar to the particular
regularization scheme we used.


\end{document}